\documentclass[10pt,journal,compsoc]{IEEEtran}
\usepackage{cite}
\usepackage{amsmath,amssymb,amsfonts,amstext}
\usepackage{algorithmic}
\usepackage{graphicx}
\usepackage{textcomp}
\usepackage[ruled,vlined]{algorithm2e}
\usepackage{booktabs,ragged2e}
\usepackage{caption}
\usepackage{subcaption}
\usepackage[flushleft]{threeparttable}
\usepackage{tikz}
  \usetikzlibrary{spy}
  \usetikzlibrary{shapes}
  \usetikzlibrary{shapes.geometric}  
  \usetikzlibrary{arrows}
  \usetikzlibrary{calc}
  \tikzstyle{ResBlock} = [draw, rectangle, minimum height=3em, 
  fill= blue, draw= blue, text=white, font=\Large, align=center, rotate=90, opacity=0.75]
  \tikzstyle{ResNode} = [draw, rectangle, minimum height=2em, minimum width=2em, 
  fill= blue, draw= black, opacity=1]
  \tikzstyle{ReLu} = [draw, ellipse, minimum height=2em, minimum width=4em, 
  fill= gray, draw= gray, text=white, rotate=90, opacity=0.75]
\usepackage{todonotes}

\newcommand{\sty}{\mathsf{y}}
\newcommand{\sta}{\mathsf{a}}
\newcommand{\argmin}{\operatorname*{arg\,min}}

\def\BibTeX{{\rm B\kern-.05em{\sc i\kern-.025em b}\kern-.08em
    T\kern-.1667em\lower.7ex\hbox{E}\kern-.125emX}}
\begin{document}
\title{Deep Learning for Material Decomposition in Photon-Counting CT}
\author{Alma Eguizabal, Ozan Öktem, and Mats U. Persson

\thanks{This study was funded by Swedish Foundation of Strategic Research under Grant AM13-0049, by MedtechLabs, by the European Union’s Horizon 2020 research and innovation programme under the Marie Sklodowska-Curie grant agreement No. 795747 and by The Swedish Research Council under grant No. 2021-05103. M. U. Persson and A. Eguizabal disclose research collaboration with GE Healthcare.}
\thanks{A. Eguizabal and O. Öktem are with the
Department of Mathematics, KTH Royal Institute of Technology, Stockholm, Sweden (e-mail: almaeg@kth.se). M. U. Persson is with the Department of Physics, KTH Royal Institute of Technology, Stockholm, Sweden}}

\maketitle

\begin{abstract}
 Photon-counting CT (PCCT) offers improved diagnostic performance through better spatial and energy resolution, but developing high-quality image reconstruction methods that can deal with these large datasets is challenging.
 Model-based solutions incorporate models of the physical acquisition in order to reconstruct more accurate images, but are dependent on an accurate forward operator and present difficulties with finding good regularization. Another approach is deep-learning reconstruction, which has shown great promise in CT. However, fully data-driven solutions typically need large amounts of training data and lack interpretability. To combine the benefits of both methods, while minimizing their respective drawbacks, it is desirable to develop reconstruction algorithms that combine both model-based and data-driven approaches. In this work, we present a novel deep-learning solution for material decomposition in PCCT, based on an unrolled/unfolded iterative network. We evaluate two cases: a learned post-processing, which implicitly utilizes model knowledge, and a learned gradient-descent, which has explicit model-based components in the architecture. With our proposed techniques, we solve a challenging PCCT simulation case: three-material decomposition in abdomen imaging with low dose, iodine contrast, and a very small training sample support. In this scenario, our approach outperforms a maximum likelihood estimation, a variational method, as well as a fully-learned network.
\end{abstract}


\section{Introduction}
\label{sec:introduction}
\IEEEPARstart{P}{hoton} Counting Computed Tomography (PCCT) is an emerging tomographic detector technology that is expected to revolutionize medical imaging \cite{Danielsson20}. These detectors have higher spatial resolution, as well as the energy (spectral) resolution that is necessary for truly using spectral CT in clinical image-guided decision making. 
There are, however, several computational challenges associated with attempts at taking full advantage of tomographic data collected by detectors in PCCT, since the higher resolution and multi-channel acquisition leads to increased complexity and larger dataset sizes compared to conventional CT.

Deep learning offers novel means to address these challenges.
Numerous studies have shown that deep learning-based image reconstruction and denoising for CT has the potential to give significant improvement in image quality \cite{Wang18}. 
Moreover, reconstruction methods that use domain adapted deep neural networks, which incorporate a handcrafted physics model \cite{Adler17PD}, have vastly better generalisation properties than approaches that rely on post-processing or on generic deep neural networks \cite{Arridge19}. 


\subsection{Image reconstruction in spectral CT}
Spectral CT techniques, including PCCT, have the ability to infer information about the atomic composition of imaged objects from measurements of the energy distribution of transmitted x-ray photons. This makes it possible to perform material decomposition, a mathematical process whereby the energy dependence of the linear attenuation coefficient is estimated in every image voxel, under the assumption that it can be expressed as a linear combination of the attenuation coefficients of a small number of basis materials. There are two paradigms of reconstruction techniques that incorporate material decomposition: one-step and two-step methods \cite{Mory18}. 
A one-step method attempts at jointly performing material decomposition and image reconstruction. Heuristic algorithms are used for this purpose, e.g., with a non-convex primal-dual method \cite{Chen21}. However, one-step methods are generally very slow and very complex to solve. Consequently, the typical approach consists of a two-step method: first performing a material decomposition in the projection domain (i.e., before image reconstruction) and then solving a tomographic reconstruction applied independently to each of the material projections decomposed from the first step.

In this work we will consider a two-step material decomposition approach and focus on the first step: the material decomposition in the projection domain. In this procedure material concentration signals are obtained from the measured spectral CT data by solving a non-linear inverse problem. The most accepted solution to this inverse problem is a maximum-likelihood (ML) estimate \cite{Roessl07}  \cite{Gronberg20} \cite{Si18}. This is a model-based method, i.e., it relies on an accurately defined forward model and does not use training data. This forward model depends on the energy distribution emitted by the source and the spectral response of the detector, and it is usually obtained by a calibration process that may introduce imprecision. Furthermore, the optimization solvers may be slow or sensitive to noise due to the ill-posedness and non-linearity of the problem \cite{Abascal18}.
Several solutions have been proposed in the last few years to improve the material decomposition. In \cite{Alvarez11} the author uses a linear approximation of the forward model to accelerate the calculations, while the authors in \cite{Ducros17} propose a regularized variational method based on a least-squares approximation to the Poisson likelihood and a regularization term. However, how to choose a good regularization functional and its parameters remains an open question. 
\subsection{Deep learning for inverse problems in tomography}
One important method of reducing noise in inverse problems is by taking advantage of prior information. The latest and very promising trend is to consider deep learning to add this prior information from training data, an approach that has demonstrated great success in tomographic reconstruction \cite{Wang20}. This step can be applied either to sinograms \cite{Lee19}  or  reconstructed images \cite{Chen17}. 

A very promising trend in the inverse problems community, and in particular for reconstruction algorithms, is to combine model- and data-driven concepts \cite{Arridge19}. Whereas a model-based approach is based on a forward model and statistical properties but does not need training data, fully data-driven (fully-learned) approaches consist of 'black box' neural networks that simply map an input to an output. 
Recent research emphasizes that most benefit is reached when combining these two paradigms. There are many different proposals to achieve this. In \cite{Shlezinger19} they propose to add neural-network blocks in a model-based algorithm to learn only specific steps that could take advantage from training data, as for instance, the estimate of a model parameter. Another interesting approach is to start from a model-based optimization problem, i.e., an objective function to minimize that is based on data fidelity and regularization terms, and include deep-learning based components inside the iterative solving method. Such deep networks can accelerate the convergence of the iterative process, as proposed in \cite{Banert20}. They can also simplify the optimization process, as suggested in \cite{Lunz21}, where a neural network performs a forward operator correction to avoid using a complex operator in the inverse problem.
Deep learning tools have also been proposed to facilitate regularization in a model-based optimization. For instance, a hand-crafted regularization term can be substituted by a network that acquires information from training data, as presented in \cite{Ulyanov20} and \cite{Baguer20}, where the authors use a generative network as regularization term. In \cite{Muk21} the authors propose to learn a convex regularizer with deep learning. Also, in \cite{Rudzu21} a set of dictionaries for a sparse optimization is learned. 

A different and very successful technique to aim for a good model-based data-driven balance is the deep algorithm unrolling \cite{Monga21}, which is what we have considered in this paper. The unrolling or unfolding framework provides a connection between deep networks and the iterative algorithms used in the model-based solutions: each of the iterating steps is substituted by a neural network that mimics the update, that is, the iterative solver is unrolled into learnable network blocks. This framework provides conceptual interpretation to deep learning, results in more powerful and robust networks, and also allows for smaller training datasets \cite{Monga21}. 
The unrolling framework was first successfully proposed in \cite{Gregor10} for sparse coding, and for the last few years has gained attention from the medical imaging community. In \cite{Adler17GD} an unrolling of a gradient-descent algorithm is proposed to solve inverse problems such as low-dose CT reconstruction. Here, the authors propose a gradient-like iterative scheme that is both model based (because it depends on the forward operator, the statistical noise model and a prior-based regularizer), and data-driven (since the gradient updates are learned using convolutional neural networks). The same authors also present a more advanced unrolling in \cite{Adler17PD}, where they apply the same framework with an unrolled Primal-Dual Hybrid Gradient and outperform the classical reconstruction methods. In \cite{Hauptmann20} a multi-scaled unrolling is proposed to deal with the extensive memory usage in CT reconstruction, reporting also that an unrolled method is more robust to previously unseen noise properties than a data-driven network. The authors in \cite{Boink19} use the unrolling framework in Photoacustic reconstruction and achieve higher quality in their results, as well as better robustness to variations in the images.
 
 




In photon-counting CT, deep learning-based approaches have been applied for denoising \cite{Mechlem17} and artifact reduction \cite{Fang20}. In particular, deep learning has been proposed for solving the material composition problem in PCCT. In \cite{Abascal20} and \cite{Holbrook21}, the authors use the well-known U-Net architecture to solve this problem. This architecture is agnostic to the physics and statistics of the problem, making this approach a fully-learned solution that maps the energy-bin sinograms to the material bases.  


\subsection{Contribution and paper overview}
In this paper we present a novel deep learning technique to solve the ill-posed material decomposition problem in spectral CT: a deep unrolled/iterative network.
We choose this approach because of its potential advantages with respect to the state-of-the-art:
\begin{itemize}
    \item Combining model- and and data- driven methods can give a good trade-off between the advantages and disadvantages of each  and potentially provide better accuracy than each of these strategies can provide individually.
    \item Incorporating model-based components into the network architecture can provide robustness leading to a network that needs fewer parameters and smaller training sets.
\end{itemize}
To evaluate the capabilities of the proposed technique, we investigate it for a challenging case: three-material decomposition into soft tissue, bone and iodine for a low-dose abdominal acquisition and a small training sample regime. This is a clinically important but difficult problem due to the location of the iodine K-edge near the lower end of the diagnostic range. The technique is evaluated for a simulated dataset generated from a realistic model of a silicon-based PCCT system with eight energy bins, incorporating spectral distortion due to non-ideal effects such as Compton scatter.
 Preliminary investigations of the proposed method have been presented previously in \cite{Eguizabal21} and \cite{Eguizabal21_2}.






\section{Material decomposition in spectral CT} \label{sec:MatDec}
A photon-counting detector consists in a multi-bin system with $B > 2$ energy bins. Each of the bins, $j=1,\dots,B$, registers the projected energy from different sections of the energy spectrum, and therefore has a different energy response. Let us consider (cross section) images in 2-D and a detector model that is uniform along its length. The expected value of the number of photon counts in bin $j$ at projection line $\ell$ follows the polychromatic Beer-Lambert law, given as
\begin{equation}\label{eq:BeerLam}
\lambda_j^{\ell} = \int _0 ^{\infty} \omega_j(E) \exp \Bigl(-\int_{\ell} \mu(s; E) ds\Bigr)dE,
\end{equation} 
where $\mu(s;E)$ is the attenuation coefficient with $s$ denoting a spatial position and $E$ the energy, and $\omega_j(E)$ models an ensemble of effects: an energy dependent X-ray source, the detector efficiency, and the energy response in bin $j$ \cite{Roessl07}. 
Note here that the integration over $\ell$ is a line integral of the integrand taken along the line $\ell$.
We assume that the attenuation coefficient can be linearly decomposed into $M$ components, that is,
\begin{equation}
\mu(s; E) \approx \sum_{m=1}^{M} \alpha_{m}(s)\tau_{m}(E),
\end{equation}
where $M$ is the number of  materials. The decomposition is typically considered in the sinogram domain (before reconstruction). Therefore, the target variable is the line integral of the materials, defined as 
\begin{equation} \label{BeerLamDec}
  a_{m}(\ell) :=
  \mathcal{T}(\alpha_{m})(\ell)
  :=\int_{\ell} \alpha_{m}(s) d s, 
\end{equation}
where $\mathcal{T}$ is the ray-transform operator.


\subsection{The inverse problem}\label{sec:InvProb}
The material decomposition is a non-linear inverse problem that consists in mapping the measured photon counts from the multi-bin detector to the material line integrals as defined in eq.~\eqref{BeerLamDec}. Let us define the Hilbert spaces $X$ for the material variables (solution space) and $Y$ as the photon counts (measurement space). The solution variable $a \in X$ is a vector containing the components of every material, i.e., $a(\ell) := [a_{1}(\ell), \ldots, a_{M}(\ell)]$ and for simplicity, we will henceforth omit the line $\ell$ from the  notation. 

The model for how a material sinogram $a \in X$ gives rise to a sinogram $y \in Y$ in absence of observation noise can now be modelled as an operator $\mathcal{F} \colon X \to Y$ that is given as
\begin{equation}\label{eq:FwdOp}
\mathcal{F}(a) := [\lambda_1(a), \lambda_2(a), \dots, \lambda_B(a)],
\end{equation}
where 
\begin{equation}
  \lambda_j(a) :=  \int_{0}^{\infty} \omega_j(E) \exp \Bigl(-\sum_{m=1}^{M} a_{m} \tau_{m}(E)\Bigr) dE,
\end{equation}
considers the poly-chromatic Beer-Lambert law on each component.
A common statistical model for the observation noise is to assume Poisson noise for the photon counts, where $j$:th energy component in measured data is Poisson distributed with mean $\lambda_{j}(a)$.
Then, measured data $y \in Y$ can be seen as a single sample of a random variable $\sty$ of the form
\begin{equation}\label{eq:Data}
\sty := \bigl[ \sty_{1}, \ldots, \sty_{B} \bigr]^{\top}
\quad\text{with $\sty_{j} \sim \text {Poisson}\bigl(\lambda_{j}(a)\bigr)$.}
\end{equation}
 Finally, based on the above, we can formalise material decomposition as the inverse problem of recovering the unknown material sinogram $a \in X$ from measured data $y \in Y$ that is a sample of the random variable $\sty$ in eq.~\eqref{eq:Data}. Note that $\mathcal{F} \colon X \to Y$ in eq.~\eqref{eq:FwdOp} serves as forward operator.


\subsection{Model-based ML estimate}
Most accepted methods to solve the material decomposition are model-based. The solution to this non-linear inverse problem is often interpreted as an ML estimation of $a$ \cite{Alvarez11} \cite{Gronberg20}. 
This estimation consists, after applying $\log$ and simplifying the Poisson likelihood expression, in minimizing the negative data  log-likelihood function $a \mapsto \mathcal{L}\bigl(\mathcal{F}(a),y\bigr)$ given as 
\begin{equation}\label{eq:likelihoodOrig}
    \mathcal{L}\bigl(\mathcal{F}(a),y\bigr) :=\sum_{j=1}^{B}\Bigl(\lambda_{j}(a)-y_{j} \log\bigl(\lambda_{j}(a)\bigr)\Bigr).
\end{equation}
%
However, the above approach for solving the material decomposition problem is \emph{ill-posed} meaning that, there can be multiple solutions (non-uniqueness) and small perturbation to data $y$ may result in a large perturbation to the minimizer (instability).
As a consequence, an iterative scheme for computing a minimizer to eq.~\eqref{eq:likelihoodOrig} may not converge to a stable solution. This is e.g.\@ the case when the non-convex minimization is solved using a scheme from a convex solver \cite{Ducros17}.

To address the above issues, and in particular the lack of stability, one needs to adopt a regularization strategy.
A simple and straightforward regularization is to add a non-negative constraint to the minimization \cite{Gronberg20}, thus leading to a constrained ML estimation:
\begin{equation}\label{eq:ML}
\begin{split}
& \min_{a} \sum_{j=1}^{B}\Bigl(\lambda_{j}(a)-y_{j} \log \bigl(\lambda_{j}(a)\bigr)\Bigr) \\[0.25em]
& \text{subject to $a_{i} \geq 0$ for all $i=1, \ldots, M$.}
\end{split} 
\end{equation} 
Projected gradient-descent is a basic iterative scheme for solving eq.~\eqref{eq:ML}. 
The iterates consist of first updating the solution according to a gradient director and then, performing an additional step by projecting the solution onto the feasibility set $Q$ defined by the inequality side-conditions in eq.~\eqref{eq:ML} (see Algorithm~\ref{alg:projGD} for further details).
\begin{algorithm}[ht]
\SetAlgoLined
 initialization $a^{0} \in Q$
 
 \For{$n = 1, \dots, N$}{
 $a^{n+1} \gets {a}^{n}-\gamma \nabla\mathcal{L}\left({a}^{n}\right)$\\
 $a^{n+1} = \operatorname{Proj}_Q(a^{(n)})$
 }
\algorithmicreturn{} $a^{N}$
\caption{Projected gradient-descent}
\label{alg:projGD}
\end{algorithm}

Such a scheme is not necessarily the best approach for solving the constrained ML estimation problem in eq.~\eqref{eq:ML}. 
However, it will serve as a blue-print for the unrolling method that we use in the following sections to define a domain adapted neural network architecture for material decomposition.
This neural network is then trained in a supervised manner to solve the inverse problem arising in material decomposition.





\section{Proposed Deep Learning solutions} \label{sec:PropDL}
Material decomposition is a non-linear ill-posed inverse problem. 
Most model based approaches will involve a data fidelity term, which is given as an appropriate affine transformation of the negative data log-likelihood $a \mapsto \mathcal{L}\bigl(\mathcal{F}(a),y\bigr)$ in eq.~\eqref{eq:likelihoodOrig}.
As an example, variational models seek to minimize an objective that is the linear combination of the aforementioned data fidelity and a regularizer, the latter represented by a hand-crafted functional that enforces stability. 

A challenge in variational models is to select an appropriate regularizer, another is to manage the computational complexity. 
For the latter, note that the objective to be minimized is non-convex due to the non-convexity of the data fidelity term. 
Hence, iterative schemes for minimizing the objective need to be properly initialized and they tend to be very time consuming. 
This becomes even more challenging in cases where the regularizer is complicated, e.g., popular sparsity promoting regularizers from compressed sensing are non-differentiable.
A final challenge with all model-based approaches, and in particular variational models, is that they assume a sufficiently accurate forward operator $\mathcal{F} \colon X \to Y$ in eq.~\eqref{eq:FwdOp}. This can be challenging since including the variety of physical phenomena necessary for sufficient accuracy (calibration corrections, pixel pile-up, detector cross-talk, \ldots)  results in a forward operator that is computationally demanding. 

Deep learning- based approaches offer a possibility to address many of the drawbacks outlined above that come with model-based approaches. 
The idea is to represent a material decomposition method by a deep neural network $\mathcal{R}_{\theta} \colon Y \to X$, which then is trained against example data.
The latter amounts to learning the (high dimensional) parameter $\widehat{\theta}$ from training data by setting up an appropriate statistical learning problem. 
The corresponding solution method is then given as
$\mathcal{R}_{\widehat{\theta}} \colon Y \to X$ (note here that the learning produces a solution method, not a specific solution).

The specific parametrization of solution methods $\mathcal{R}_{\theta}$ is dictated by the choice of deep neural network architecture.
In the fully-learned setting, the architecture will not account for an explicit hand-crafted (aka non-learned)  forward operator that encodes the physics of how training data is generated.
This has some downsides when the trained network  $\mathcal{R}_{\widehat{\theta}} \colon Y \to X$ is suppose to solve an ill-posed and high dimensional inverse problem, like the material decomposition problem.
For good results one needs to learn $\widehat{\theta}$ from a vast amount of training data. 
Such data are difficult to come by in medical imaging and especially so for emerging imaging technologies like PCCT.
Next, even with large amounts of training data, the resulting learned solution operator $\mathcal{R}_{\widehat{\theta}}$ generalizes poorly.

One way to address the above issues is to better adapt the deep neural network architecture to the specific problem at hand. 
One such domain adaptation is to account for the fact that a trained solution operator is a (regularized) approximation to the inverse of the forward operator, i.e., $\mathcal{R}_{\widehat{\theta}} \approx \mathcal{F}^{-1}$.
We aim to explore two ``physics-aware'' deep learning approaches of this type, both based on unrolling a suitable iterative scheme.
The first approach considers a neural network architecture given by considering a learned update to a gradient-descent scheme as in \cite{Eguizabal21}.
The second approach includes a more explicit use of the physics with a learned update function that also casts the value of the gradient of the likelihood cost used in the model-based approach. 

Both the above deep neural networks could be computationally demanding to train, but once trained they are fast to apply in runt-time. Next, besides the choice of architecture, a user only needs to provide training data. In particular, there is no need to select an explicit regularizer and set values for its hyper-parameters. 



\subsection{Learned post-processing}\label{sec:UnrollDenoiser}
The idea here is to define the solution operator as applying a learned post-processing to an initial material decomposition, i.e., $\mathcal{R}_{\theta}(y) := \Gamma_{\theta}\bigl(\mathcal{R}_{ML}(y)\bigr)$ with $\theta=(\theta^{1},\ldots,\theta^{N})$ where $\mathcal{R}_{ML} \colon Y \to X$ is an initial decomposition method, say one based on ML as in $\mathcal{R}_{ML}(y) := \hat{a}_{ML}(y)$, and $\Gamma_{\theta}(a) = a^N$ is a learned post-processing that is given as
\begin{equation}
\begin{split}
a^{0} &= a \\
a^{n} &= a^{n-1}-\Psi_{\theta^{n}}(a^{n-1})
\end{split}
\quad\text{for $n=1,\ldots, N$.}
\end{equation}
Here, $\Psi_{\theta^{n}}$ is the $n$:th residual block with $\theta^{n}$ representing the convolution filters, so $\Gamma_{\theta} \colon X \to X$ is a residual deep neural network.
Note that in the above, the architecture for $\Gamma_{\theta}$ \emph{does not} account for the forward operator and also does not assume any model information in its internal structure. 
It is merely a denoiser of an initial solution.
Hence, the physics and statistical models are only implicitly accounted by $\mathcal{R}_{ML}$.
\begin{figure}[h!]
  \centering
   \begin{subfigure}[b]{0.39\textwidth}
         \centering
        \includegraphics[width=1\textwidth]{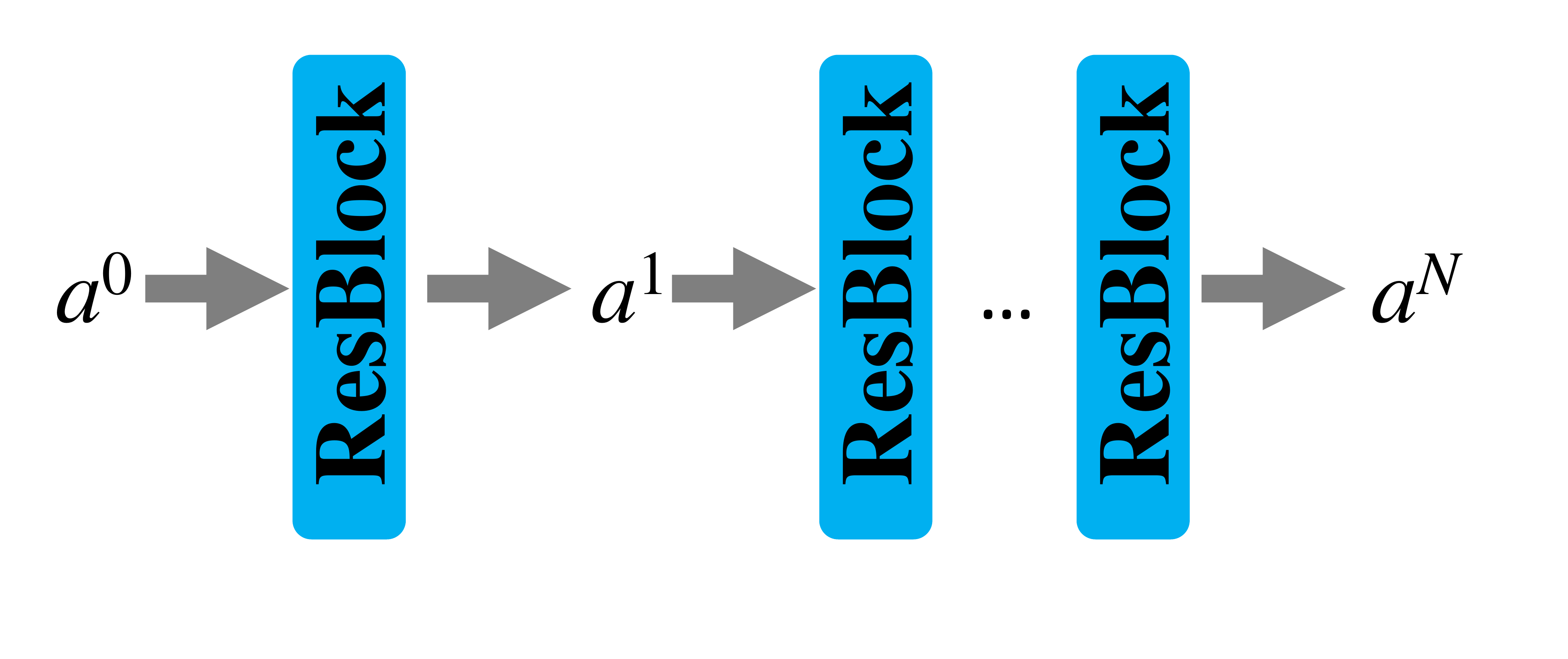}
         \caption{Learned post-processing}
    \end{subfigure}  
    \par\bigskip 
    \begin{subfigure}[b]{0.39\textwidth}
         \center
        \includegraphics[width=1\textwidth]{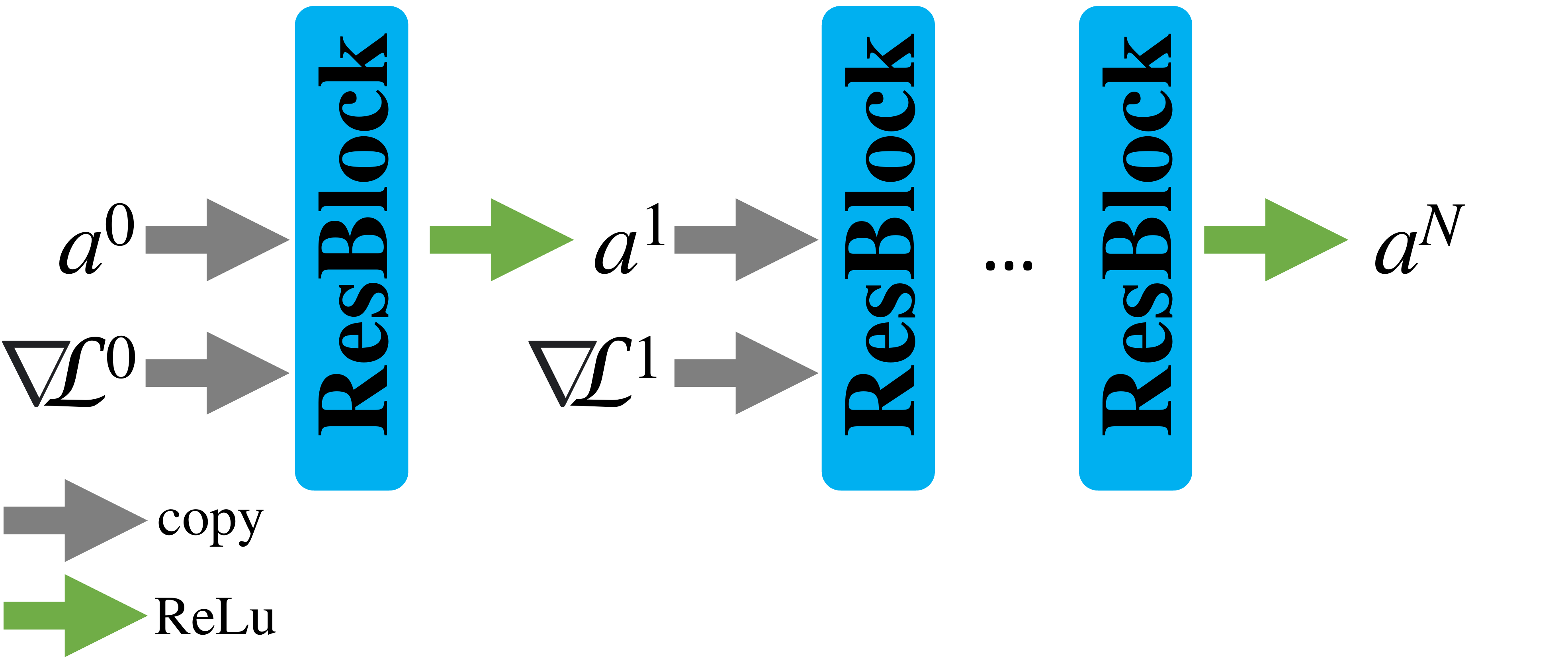}
         \caption{Learned gradient-descent}
    \end{subfigure}
   \caption{Proposed deep learning solutions. (a) A post-processing technique that mimics updates with residual blocks. (b) An unrolled gradient-descent scheme, which also incorporates the gradient of the likelihoodin each block. Both architectures rely on stacking residual blocks of the form in Fig.~\ref{fig:archis2}.
}
  \label{fig:archis}
\end{figure}

\begin{figure}[h!]
  \centering
  \includegraphics[width=0.39\textwidth]{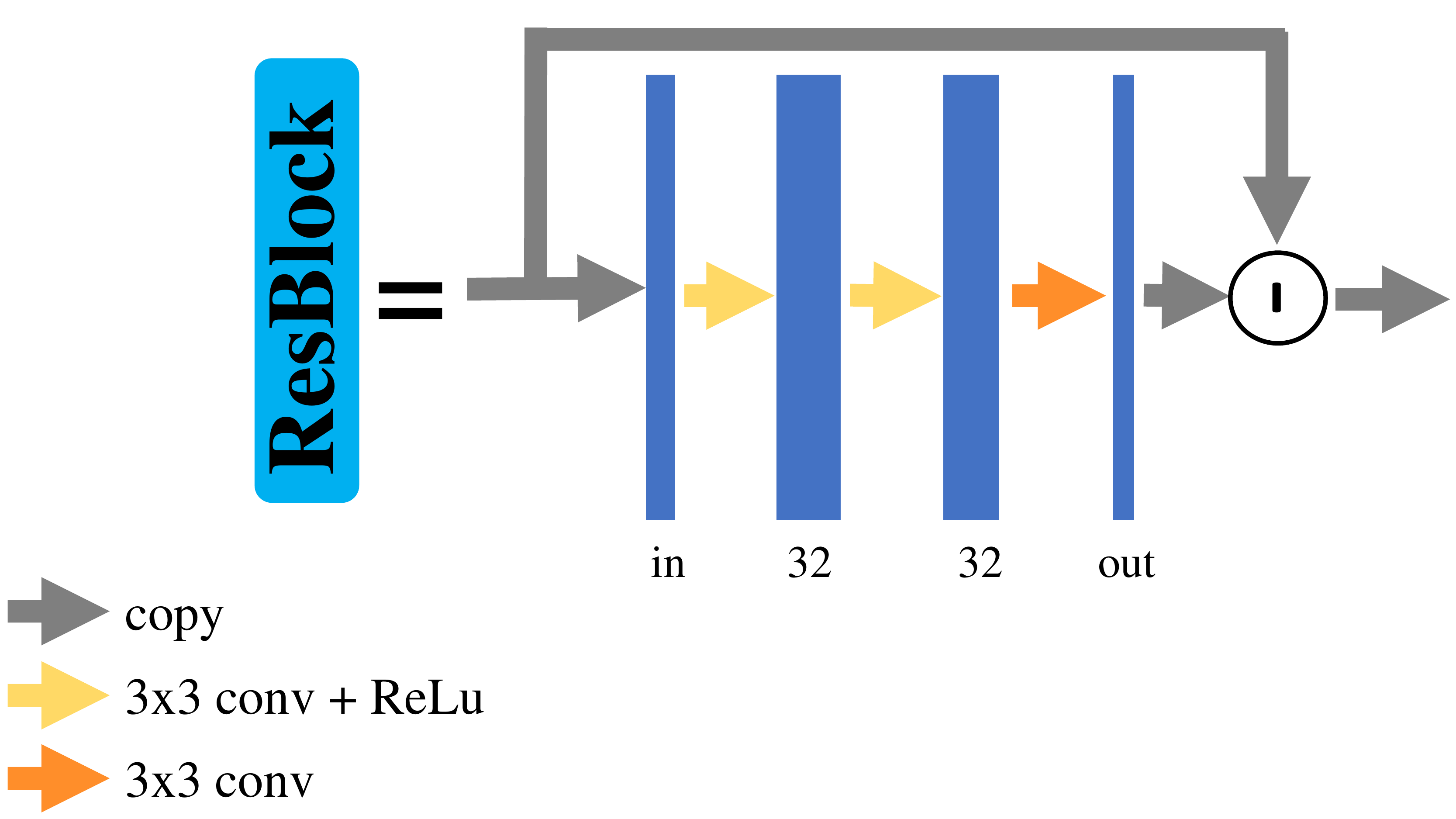}
  \caption{Residual blocks that conform the proposed solutions.}
  \label{fig:archis2}
\end{figure}

\begin{algorithm}
\SetAlgoLined
 initialization $a^{0} \in Q$
 \For{$n = 1, \dots, N$}{
 $a^{n+1}=a^{n}-\Psi_{\theta^{n}}\left(a^{n}\right)$\\
 }
\algorithmicreturn{} $a^{N}$
 \caption{Learned update function} \label{alg:ResBl}
\end{algorithm}

\subsection{Learned gradient-descent}\label{sec:LearnedGrad}
This proposed architecture is derived by unrolling a gradient-descent  scheme originally designed to minimize the  handcrafted data fidelity in eq.~\eqref{eq:likelihoodOrig}, i.e., a scheme of the form 
\[ 
  a^{n+1} := a^{n}-\gamma \nabla\mathcal{L}(a^{n}, y).
\]
Truncating the above scheme after $N$ iterates and replacing the handcrafted updates with learned ones results in an operator $\mathcal{R}_{\theta}(y) = a^N$ where $\theta=(\theta^{1},\ldots,\theta^{N})$ and 
\begin{equation}
a^{n} = a^{n-1} 
  - \Psi_{\theta^{n}}\bigl(a^{n-1},\nabla\mathcal{L}(a^{n-1}, y) \bigr)
  \quad\text{for $n=1,\ldots,N$.}
\end{equation}
See also Algorithm~\ref{alg:uGD}. Note here that each $\Psi_{\theta^{n}}$ is a residual block operation that is learned during training, whereas the gradient $\nabla \mathcal{L}(a^{n}, y)$ is handcrafted as 
\begin{equation}
\frac{\partial \mathcal{L}}{\partial a_m}(a) =  
\sum_{j=1}^B \Bigl( \frac{y_j}{\lambda_j}-1 \Bigr)
\int_{0}^{\infty}\!\!\! \tau_{m} w_j  \exp \Bigl(-\sum_{i=1}^{M} a_{i} \tau_{i}(E)\Bigr) dE.
\end{equation}
Here, $\mathcal{R}_{\theta} \colon Y \to X$ is a deep neural network with an architecture that incorporates the information from the physics of how data is generated as well as a statistical models for the  observation errors.

The above neural network architecture for material decomposition shares many similarities with the one outlined in Section~\ref{sec:UnrollDenoiser} that is essentially a denoise.
In fact, their structure is identical, except for the extra ReLus and the gradient of the data log likelihood, which can be viewed as an additional elements in the learned gradient-descent. This difference is also depicted in Fig.~\ref{fig:archis}. 
\begin{algorithm}
\SetAlgoLined
 initialization $a^{0} \in Q$
 \For{$n = 1, \dots, N$}{
 $a^{n+1}=a^{n}-\Psi_{\theta^{n}}\bigl(a^{n}, \nabla\mathcal{L}(a^{n}, y)\bigr)$
 }
\algorithmicreturn{} $a^{N}$

 \caption{Learned gradient-descent}
 \label{alg:uGD}
\end{algorithm}


To avoid numerical instabilities (for near zero values) during the training, we consider approximating the the Poisson likelihood functional in eq.~\eqref{eq:likelihoodOrig} with a squared weighted $\ell^2$-norm \cite{Tianfang04}.
To this aim, we calculate the log of the photon counts with $z = [z_1, \dots, z_B]$ and expected counts with $\gamma = [\gamma_0, \dots, \gamma_B]$, being each element defined as
\begin{equation}
z_j = -\log(y_j / N_{0j} )
\quad\text{and}\quad
\gamma_j = -\log(\lambda_j / N_{0j}),
\end{equation}
where $N_{0j}$ are the photon counts in the detector on air, for each $j$ energy bin.
The new log likelihood functional is then the squared weighted L2-norm:
\begin{equation}
\mathcal{L}_{LS}(z,\gamma) = (z-\gamma) \cdot \Sigma \cdot (z-\gamma)^{\top},
\end{equation}
where $\Sigma := \operatorname{diag}(1/y_j)$.
Thus, the gradient of the data fidelity can be approximated by the following:
\begin{equation}
\frac{\partial \mathcal{L}_{LS}}{\partial a_m} =  
\sum_{j=1}^B \frac{1}{y_j}\frac{\gamma_j-z_j}{\lambda_j}
\int_{0}^{\infty}\!\!\! \tau_{m} w_j  
\exp\Bigl(-\sum_{i=1}^{M} a_{i} \tau_{i}(E)\Bigr) dE.
\end{equation}


\subsection{Statistical interpretation}
The aim here is to provide a statistical interpretation of the learned solution method $\mathcal{R}_{\widehat{\theta}} \colon Y \to X$. 
This requires one to phrase the inverse problem in Section~\ref{sec:InvProb} in a fully statistical setting following \cite[Sec.~3]{Arridge19}.

The first step is to introduce an additional $X$-valued random variable $\sta$ that generates the unobserved (true) material sinogram $a \in X$ one seeks to recover in material decomposition.
Measured data $y \in Y$ is now a sample of the conditional random variable $( \sty \mid \sta = a)$ where $a \in X$ is the aforementioned material sinogram and the random variable $\sty$ is given as in eq.~\eqref{eq:Data}.
The inverse problem is now formalized as the task of estimating $a$ by a suitable point estimator that summarizes the posterior distribution for $( \sta \mid \sty = y)$.
Examples of possible point estimators are posterior mean, median and mode.

An issue with the above approach is that the posterior distribution for $( \sta \mid \sty = y)$ is unknown.
On can now use Bayes' theorem to express this posterior in terms of the distributions for $( \sty \mid \sta = a)$ (data likelihood) and $\sta$ (prior).
The data likelihood is known from the physics model outlined in Section~\ref{sec:InvProb}, but the prior remains unknown.
A further issue with an approach that relies of Bayes' theorem is to manage the large scale nature of the computations that are involved in recovering the posterior.
If one has access to sufficient amount of supervised training data, then it is possible to address both the issue of unknown prior and computational complexity.

To see this, assume there is supervised training data in the form of input/output pairs $(a_1,y_1), \ldots, (a_m,y_m) \in X \times Y$ that are random samples of $(\sta,\sty) \sim \beta$.
Furthermore, assume we look to use this example data to learn a solution method from a fixed family $\{ \mathcal{R}_{\theta} \}_{\theta}$ of operators $\mathcal{R}_{\theta} \colon Y \to X$. 
One can then learn a solution method from training data above as $\mathcal{R}_{\widehat{\theta}} \colon Y \to X$ where $\widehat{\theta}$ is an (approximate) solution to the learning problem 
\begin{equation}\label{eq:Learning}
\widehat{\theta} \in \argmin_\theta \Bigl\{ \frac{1}{m}\sum_{i=1}^m \ell_{X}\bigl( \mathcal{R}_{\theta}(y_i),a_i \bigr) \Bigr\}.
\end{equation}
Here, $\ell_{X} \colon  X \times X \to \mathbb{R}$ is a loss-function that quantifies similarity in $X$-space.
A statistical interpretation of \emph{what} the learned solution method $\mathcal{R}_{\widehat{\theta}}$ represents comes from interpreting the objective in eq.~\eqref{eq:Learning} as an empirical counterpart to the $\beta$-expectation of $\ell_{X}\bigl( \mathcal{R}_{\theta}(\sty),\sta \bigr)$ where $(\sta,\sty) \sim \beta$. 
Hence, 
\begin{equation}\label{eq:Learning2}
\mathcal{R}_{\widehat{\theta}}
     \approx 
     \argmin_{\mathcal{R} \colon Y \to X} 
     \mathbb{E}_{(\sta,\sty) \sim \beta} 
     \Bigl[ \ell_{X}\bigl( \mathcal{R}_{\theta}(\sty),\sta \bigr) \Bigr],
\end{equation}
i.e., it approximates a Bayes estimator.
This means we seek the solution method for the inverse problem that minimizes the average error, the latter quantified by the $\ell_{X}$-loss.
If the loss is selected as a squared $\ell^2$-norm, then the right hand side of eq.~\eqref{eq:Learning2} equals the posterior mean, i.e., 
\begin{equation}\label{eq:Learning3}
\mathcal{R}_{\widehat{\theta}}(y)
     \approx \mathbb{E}\bigl[ \sta \mid \sty=y \bigr].
\end{equation}
The above statistical interpretation holds context of material decomposition when the solution method is trained against supervised data $(a_i,y_i) \in X \times Y$, which is e.g. the case in Section~\ref{sec:LearnedGrad}.
The interpretation changes somewhat if the solution operator is trained against other type of data, e.g., in Section~\ref{sec:UnrollDenoiser} we train a post-processing method $\Gamma_{\theta} \colon X \to X$ against pairs $(a_i,a^i_{ML}) \in X \times X$ where $a_i \in X$ is the ground truth material sinogram that is associated with some measured sinogram $y_i \in Y$ and $a^i_{ML} \in X$ is the corresponding ML estimate computed from $y_i$ by solving eq.~\eqref{eq:ML}.
The learned post-processing $\Gamma_{\widehat{\theta}} \colon X \to X$ is now an approximation to the posterior mean for $( \sta \mid \sta_{ML}=a_{ML})$, where $\sta_{ML}$ denotes the random variable generating ML estimates.

\section{Implementation and evaluation} \label{sec:ImpEv}
\begin{figure}[h!]
  \centering
  \includegraphics[width=.45\textwidth]{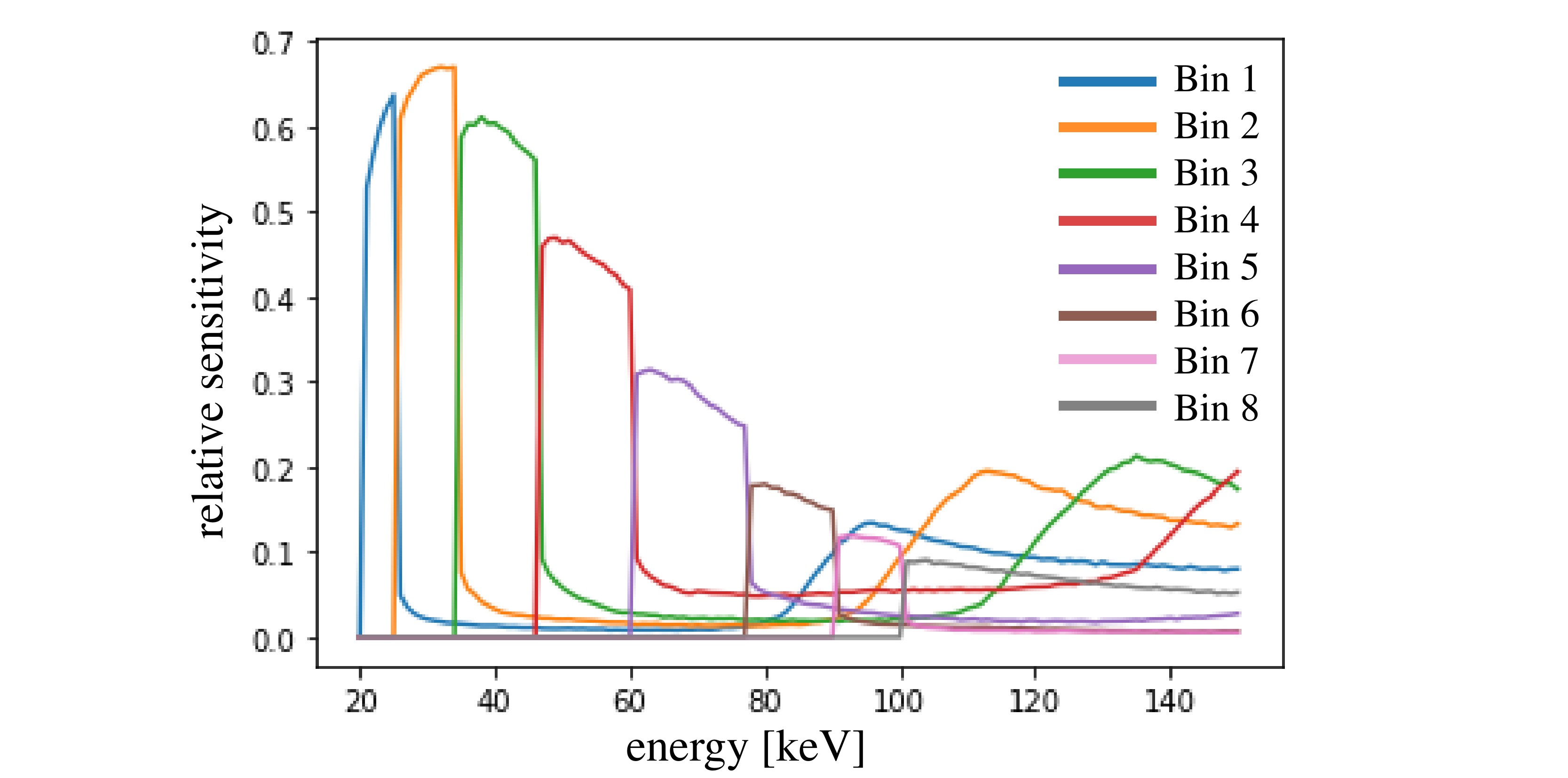}
  \caption{Silicon-based PCCT detector model, with eight-bin sensitivity and Compton scatter.}
  \label{fig:bin_sens}
\end{figure}

We have evaluated the performance of our proposed solutions in a simulation study. This simulation consists of a PCCT system with fan beam geometry, 512 detector elements, 512 angles, and an image field of view $35 \times 35$~cm. The source-to-ISO distance is 541~mm and the source-to-detector 949~mm. The detector model has eight bins, is silicon-based and considers Compton scatter \cite{Persson20}, as represented in Fig.~\ref{fig:bin_sens}. The source is 120~kVp \cite{Evans98} and has a 9.3~mm of Aluminum flat filtration. We have used ODL library \cite{Adler17ODL} to simulate the forward operator, geometry and ray transformations.

We aim to solve a challenging imaging case: abdominal imaging with iodine contrast, relative low radiation dose (100~mAs) \cite{Dion04}, and a three-material decomposition. In this case, the noise level is expected to be high in the resulting material sinograms if a typical model-based solution is used. We have generated the material images with the KiTS19 database \cite{Kits19}, which consists of regular CT volumes of the kidney, with iodine-injected patients with kidney tumor, and images of size $512 \times 512$. This database contains segmentation labels for the tumours, which correspond to areas of high iodine concentration. We choose a few slides per volume (per patient). First, we segment the images into three-material images, corresponding to bone, soft-tissue and iodine. Bone and soft-tissue are obtained by transforming the original HU units to material densities, and then performing a naive thresholding, where bone corresponds to the bigger densities. For the iodine mapping we use the segmentation information from the tumours that is provided in the database. 
We have assumed iodine concentrations of 5--10~mg/ml and a random texture. To fix the dose to 100~mAs we have defined $2.7 \cdot 10^5$~photons$/$(pixel view).

In order to keep the test case challenging, and assuming that in medical imaging training data is not abundant (an especially in an emerging technology such as PCCT), we have considered only 200 training samples, which is a relative small number (in opposition to our previous work in \cite{Eguizabal21} and \cite{Eguizabal21_2}). We use 100 test cases to illustrate and compute the results. To define and train the deep networks we have used PyTorch with Adam optimizer, and four NVIDIA GPU GeForce RTX 2080 Ti.

\subsection{Special test cases}
We have considered two additional scenarios, in which obtaining an accurate three-material decomposition is especially challenging. 
\subsubsection{Very small training dataset}
Let us assume that we only dispose of 50 training samples. We will evaluate the robustness against very small training sample support, which is usually a weakness of fully-learned approaches.
\subsubsection{Forward operator with calibration errors}
Let us consider that there are small errors in the calibration process that affect the accuracy of the forward operator. We will evaluate robustness against errors in the formulation of the forward model, to which the model-based methods are particularly sensitive. 
In order to simulate these errors we have assumed that the Aluminum filter length may vary $\pm 1.5$~mm and the kVp of the source may change $\pm 5\%$. 


\subsection{Competing strategies}
With the purpose of a comprehensive comparative study, we have investigated three state-of-the-art solutions to the material decomposition: two are model-based (as most solutions to this problem are) and one is fully-learned: 
\subsubsection{Model-based}
\begin{itemize}
    \item An ML estimation, as proposed by \cite{Gronberg20}, solved with a second-order method.
    \item A variational problem, where a first-order pseudo Huber penalty is used as regularization, as proposed by \cite{Abascal18}.
\end{itemize} 

\subsubsection{Fully data-driven} 
\begin{itemize}
    \item A ``black-box" U-Net, as proposed by \cite{Abascal20} and \cite{Holbrook21}.
\end{itemize}




\section{Results and discussion} \label{sec:ResDis}
In this section we present a collection of quantitative and qualitative results. Since the proposed technique computes results in sinogram domain we have evaluated most of the quality metrics in this domain. In order to also validate how the results affect the final image, we have used filtered-back projection (FBP) to study the performance in the image domain. As qualitative results, we present the resulting material-basis sinograms, as well as the virtual mono-image at 70~KeV with material overlay of the bases, which represent the bone and iodine concentration.
\subsection{Quantitative results}
The quantitative results are summarized in three tables. Table~\ref{tab:1} presents the mean square error (MSE) and Structural Similarity Index Measure (SSIM) of the model-based methods, i.e., the ML estimator (ML), and the variational method (Variational), and deep learning approaches, that is, the ``black-box" U-Net (Fully-learned), and our proposed learned post-processing (Learned Post.) and learned gradient-descent (Learned GD). 
In order to observe which material dominates the error for every strategy, we have also separated the MSE in terms of the different bases materials (channels) in Table~\ref{tab:2}. We evaluate the MSE for bone (MSE-b), soft-tissue (MSE-sf) and iodine (MSE-io) for the original test with 200 training samples, and compare it against the test with only 50 training samples. 
Finally, in Table~\ref{tab:3} we present the quality metrics in the image domain after FBP. 


\begin{table}[h]
\begin{threeparttable}
\caption{Quantitative results in terms of Mean Square Error (MSE) and Structural Similarity Index Measure (SSIM) for the initial test (200 training samples) and the special test cases of only 50 samples and calibration error (cal. error). Values correspond to the average of the three materials; param. refers to to parameters in the networks. }
\label{tab:1}
\setlength\tabcolsep{0pt} 

\begin{tabular*}{\columnwidth}%
  {@{\extracolsep{\fill}} l r r r r r r r r r r}
\toprule
      & 
     \multicolumn{2}{c}{200 samples}& 
     & \multicolumn{2}{c}{50 samples}&
     & \multicolumn{2}{c}{cal. error}\\ 
     
\cmidrule{2-9}
     & \multicolumn{1}{c}{MSE} & \multicolumn{1}{c}{SSIM}  & 
     & \multicolumn{1}{c}{MSE} & \multicolumn{1}{c}{SSIM} & & \multicolumn{1}{c}{MSE} & \multicolumn{1}{c}{SSIM} & \multicolumn{1}{c}{param.} \\
\midrule
    ML \cite{Gronberg20} &  1.41 & 0.28  &&  1.41  & 0.28  & &2.21 & 0.37 & \multicolumn{1}{c}{--} \\
\addlinespace
     Variational \cite{Ducros17} &15.83 &  0.08 &&  15.83&  0.08 &  &47.96 &0.00& \multicolumn{1}{c}{--} \\
\addlinespace
     Fully-learned  &  0.14 & 0.49 & & 1.27 & 0.09 && 0.20 & 0.45  & $3.1\cdot 10^7$ \\

\addlinespace
     Learned Post. & 0.08 & 0.32 & & 0.15 & 0.22 && 0.25 & 0.40& $1.1 \cdot 10^5$ \\

\addlinespace
    Learned GD  &  \textbf{0.06} & \textbf{0.83} & & \textbf{0.08} & \textbf{0.75} && \textbf{0.14} & \textbf{0.68} & $1.2 \cdot 10^5$\\

\bottomrule
\end{tabular*}

\smallskip
\scriptsize
\end{threeparttable}
\end{table}

\begin{table}[h]
\begin{threeparttable}
\caption{Comparison of the MSE per material channel for initial test (200 training samples) and the small training case (50 samples).}
\label{tab:2}
\setlength\tabcolsep{0pt} 

\begin{tabular*}{\columnwidth}%
  {@{\extracolsep{\fill}} l r r r r r r r r}
\toprule
      & 
     \multicolumn{3}{c}{200 samples}& 
     & \multicolumn{3}{c}{50 samples}\\ 
     
\cmidrule{2-8}
     & \multicolumn{1}{c}{MSE-b} & \multicolumn{1}{c}{MSE-sf} 
     & \multicolumn{1}{c}{MSE-io} & & \multicolumn{1}{c}{MSE-b} 
     & \multicolumn{1}{c}{MSE-sf} & \multicolumn{1}{c}{MSE-io} \\
\midrule
     ML \cite{Gronberg20} & 0.41 & 2.40 & 1.44 &&   0.41 & 2.40 & 1.44 \\
\addlinespace
     Variational \cite{Ducros17} & 2.26 & 17.78 & 27.44 &&    2.26 & 17.78 & 27.44 \\
\addlinespace
     Fully-learned  & \textbf{0.01} & 0.35 & \textbf{0.05} &&   0.11 & 3.46 & 0.24 \\

\addlinespace
     Learned Post. & 0.02 & 0.09 & 0.13 &&   0.07 & 0.18 & 0.21\\

\addlinespace
    Learned GD  &  \textbf{0.01} & \textbf{0.06} & 0.11 &&   \textbf{0.03} & \textbf{0.08} & \textbf{0.13} \\

\bottomrule
\end{tabular*}

\smallskip
\scriptsize
\end{threeparttable}
\end{table}

\begin{table}[h]
\begin{threeparttable}
\caption{Comparison in image domain after filtered back projection of peak-to-noise ratio (PSNR) and SSIM for 200 training samples and 50 training samples. Values correspond to the average of the three materials.}
\label{tab:3}
\setlength\tabcolsep{0pt} 

\begin{tabular*}{\columnwidth}%
  {@{\extracolsep{\fill}} l l l l l l l}
\toprule
      & 
     \multicolumn{2}{c}{200 samples}& 
     & \multicolumn{2}{c}{50 samples}\\ 
     
\cmidrule{2-6}
     & \multicolumn{1}{c}{PSNR} & \multicolumn{1}{c}{SSIM} 
     & & \multicolumn{1}{c}{PSNR} & \multicolumn{1}{c}{SSIM} \\
\midrule
     ML \cite{Gronberg20} &   21.31  & 0.02 &   & 
     21.31 (=) & 0.02 (=)  \\
\addlinespace
     Variational \cite{Ducros17} &  13.61 & 0.00 &    & 
     13.61 (=) & 0.00 (=)\\
\addlinespace
     Fully-learned  &   40.26 & 0.62 &    & 32.39 
     ($\downarrow$ 7.87)& 0.18 ($\downarrow$ 0.44)\\

\addlinespace
     Learned Post. &   41.33 & 0.62 &   &\textbf{40.45} 
     ($\downarrow$ 0.88)& 0.54 ($\downarrow$ 0.08)& \\

\addlinespace
    Learned GD  &    \textbf{42.12} & \textbf{0.72} &    & 40.08
    ($\downarrow$ 2.04)& \textbf{0.56} ($\downarrow$ 0.16)\\

\bottomrule
\end{tabular*}
\smallskip
\scriptsize
\end{threeparttable}
\end{table}

\subsection{Quantitative results}
\begin{figure*}
    \centering
    \includegraphics[width=.99\textwidth]{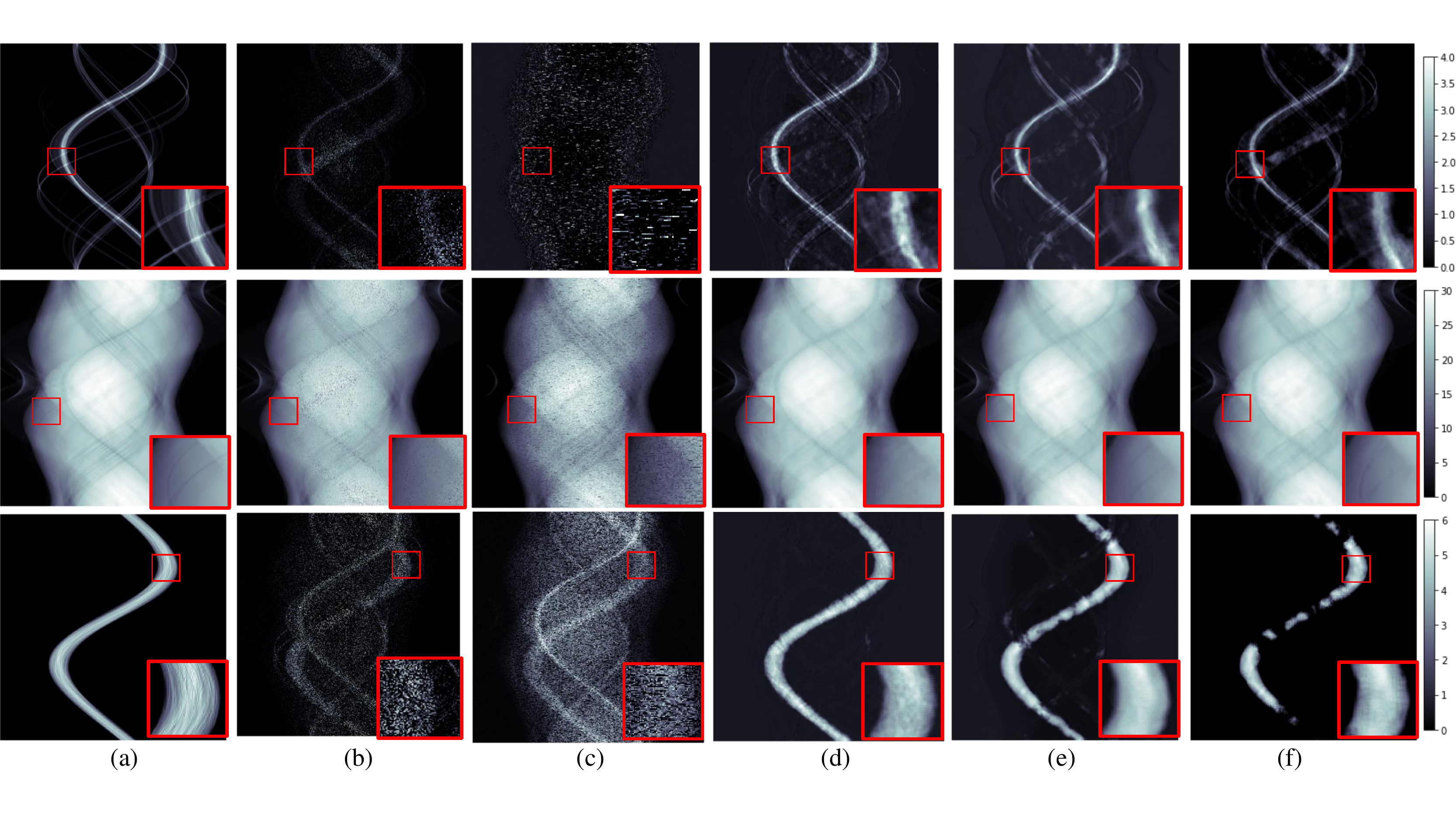}
    \caption{Resulting material-basis sinograms after a material decomposition. Each row correspond to one material (bone, soft-tissue and iodine). The columns are (a) Ground-truth (b) Model-based: ML estimation \cite{Gronberg20} (c) Model-based: Variational Method \cite{Abascal18} (d) Fully-learned U-Net (e) Proposed learned post-processing (f) Proposed learned GD. Values are in $cm$.}. 
    \label{fig:sino}
\end{figure*}

\begin{figure*}
    \includegraphics[width=0.99\textwidth]{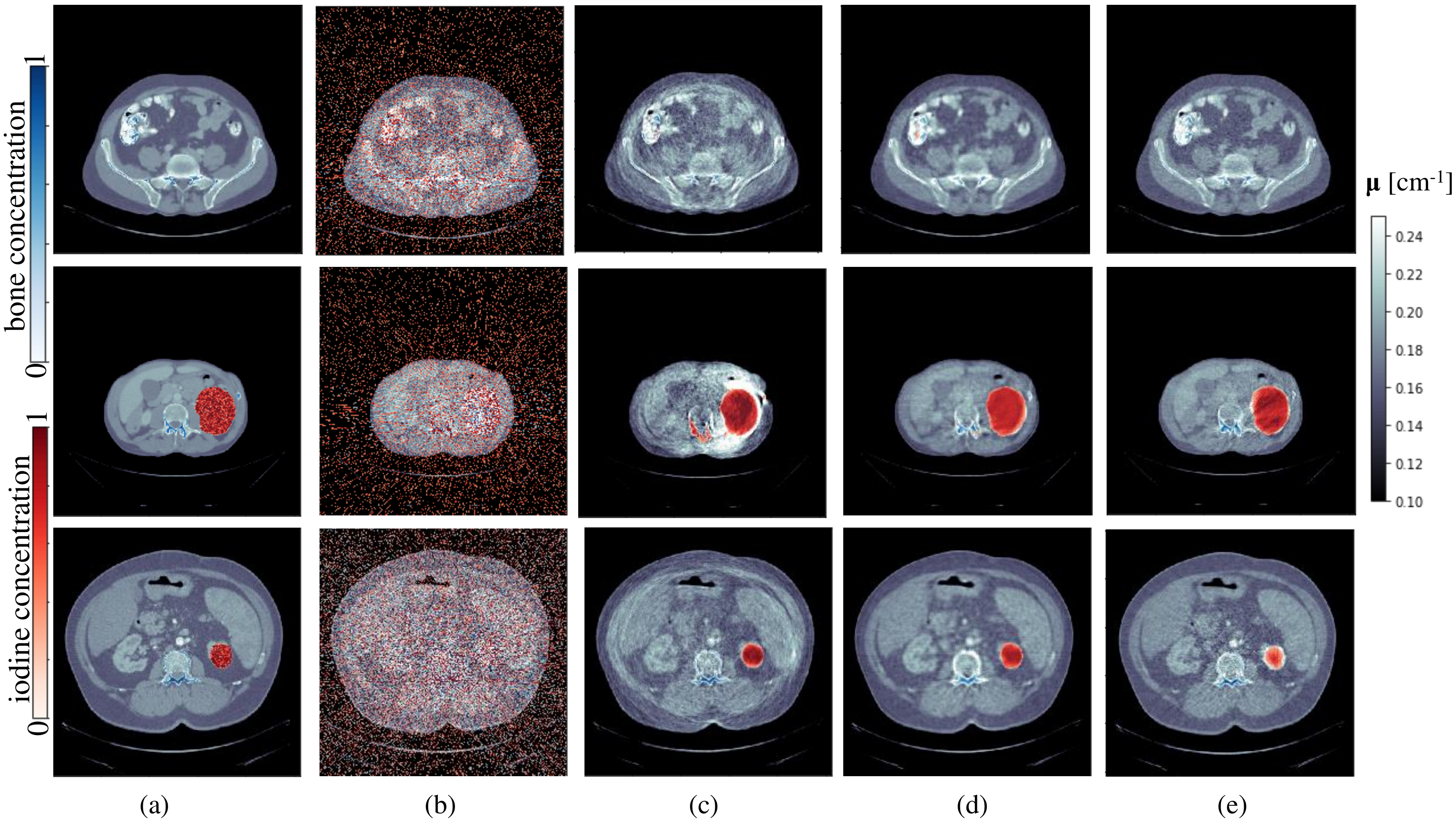}
    \caption{Virtual mono-image at 70~keV with material concentration overlay calculated from (a) The ground-truth material sinograms, (b) Model-based: the ML material estimates, (c) Fully-learned U-Net, (d) Proposed learned post-processing, (e) Proposed learned gradient-descent. Each row correspond to a different patient (sample) from KiTS19. We do not include the variational method here because its material sinograms were very noisy (as shown in Fig. \ref{fig:sino}, and after a FBP the images are very low quality, with an indistinguishable material overlay. Material values correspond to concentration (0 to 1). }
    \label{fig:image}
\end{figure*}

\begin{figure}
    \centering
   \includegraphics[width=0.49\textwidth]{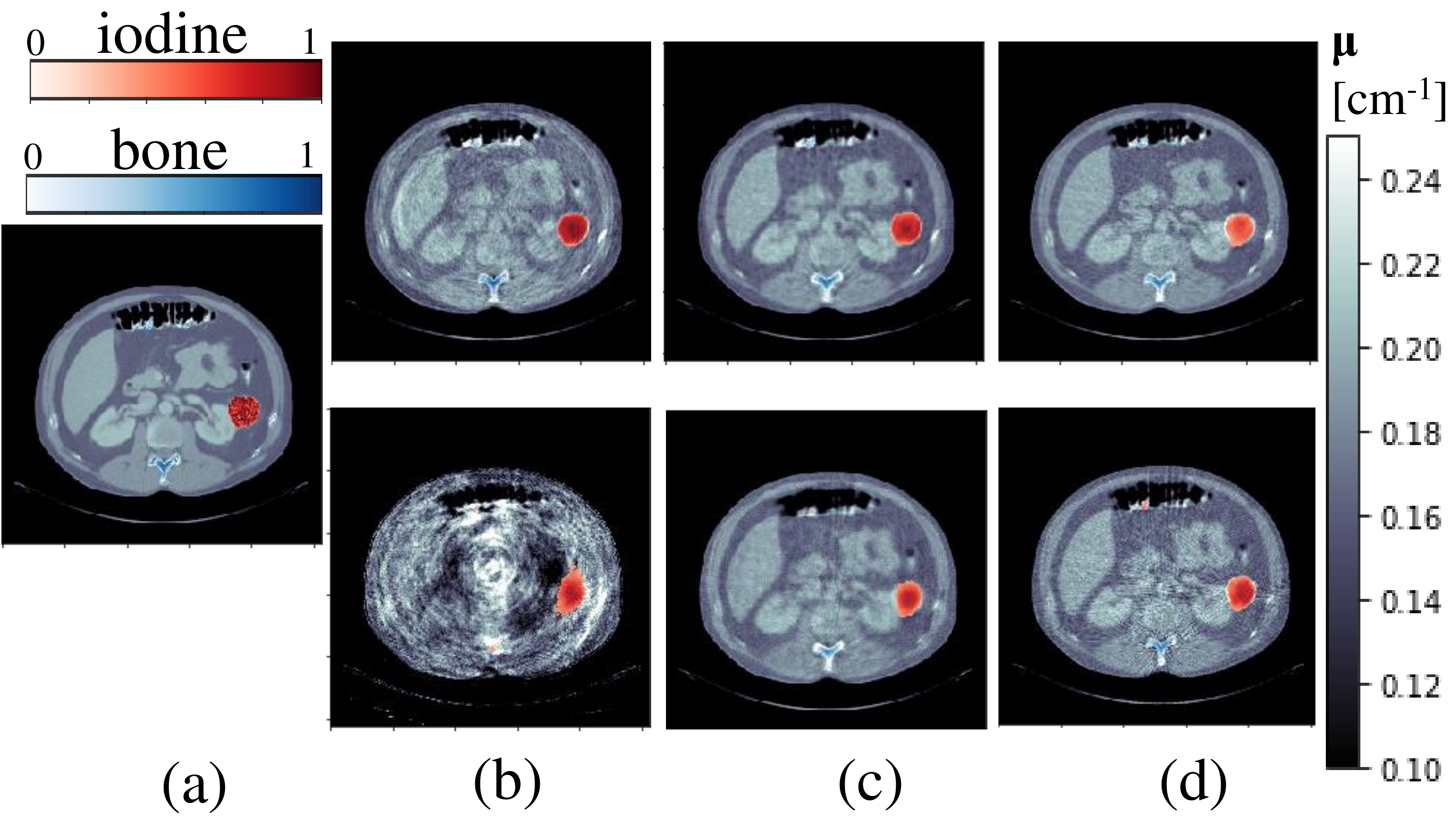}
    \caption{Virtual mono-image at 70~keV with material concentration overlay calculated from (a) The ground-truth material sinograms (b) Fully-learned U-Net, (c) Proposed learned post-processing, (d) Proposed learned gradient-descent. (In (b),(c) and (d) the upper row corresponds to 200 training samples, and lower row to 50 samples). }
    \label{fig:image_200vs50l}
\end{figure}

We present two figures to illustrate the results. First, we have depicted in Fig.~\ref{fig:sino} the basis sinograms corresponding to each material (bone, soft-tissue and iodine). 
Second, we present the virtual mono-image at 70~keV in Fig.~\ref{fig:image}, 
where each row corresponds to a different patient: first row does not have iodine, second row has a significantly big tumor and third row is a patient with a smaller tumor.
Finally, a forth patient is shown in Fig.~\ref{fig:image_200vs50l}, where in the first row we present the mono-image result of the initial test (with 200 training samples) and in the second row the very small training set case (50 samples). 



\subsection{Discussion}
Deep neural networks show excellent results: in every scenario deep learning solutions greatly overcome model-based approaches, which tend to be more noisy. As observed in Fig.~\ref{fig:sino}(b), a three-basis ML estimator is not only noisy but also contains material cross-contamination (we can see the iodine signal in the bone sinogram and vice-versa). This could be mean that, the non-negativity constraint is too simple as regularization. In addition, in the image domain, as observed in Fig.~\ref{fig:image}(b), the ML estimates result in a very poor reconstruction, especially for the bone and iodine bases. The iodine does not show localization information after FBP, which would complicate considerably any clinical application. With respect to the second model-based approach, the variational method, its performance is also very poor, as seen in Fig. \ref{fig:sino}(c), and also worst that the results reported in \cite{Ducros17}. Our selection of regularization parameters may have not been optimal, even though we followed the proposed in \cite{Spray17}, as well as tried a  parameter tuning that did not get better results. One explanation for this is that, the impact of the regularization parameters becomes more critical in a decomposition involving iodine and bone (where K-edges are very similar) instead of gadolinium, which is the contrast agent considered in \cite{Ducros17}.

Within the deep learning methods, our proposed approaches demonstrate several advantages with respect to a fully-learned U-Net, especially our learned gradient-descent. 
As summarized in Table~\ref{tab:1}, our proposed deep learning techniques have shown the best performance in terms of MSE and SSIM, besides the fact that they need smaller capacity, i.e., fewer training parameters (param. in Table~\ref{tab:1}). The unrolled gradient-descent stands out, especially for SSIM. Furthermore, when the number of training samples is reduced (50 samples), the fully-learned U-Net presents an important loss of accuracy (one order of magnitude), whereas our approaches do not suffer from such a performance drop. This is visually obvious in Fig. \ref{fig:image_200vs50l}(b), where the U-Net result is no longer acceptable. Thus, our proposed physics-informed architecture is not only allowing for fewer training parameters, but also providing more robustness in very small training-sample regimes.  
In the calibration error study (cal. error), the model-based approaches (ML and Variational) suffer from an important degradation. On the contrary, our learned gradient-descent is still accurate, even though the forward operator in its architecture is not corrected for the calibration errors either.
However, this performance could still be enhanced if the calibration scenario was explicitly considered in the network architecture. This could be done, for instance, with a dedicated learned correction for the gradient inputs in Fig.\ref{fig:archis}(b), which contain the forward model information. Such an explainable architecture modification would be, however, not straight-forward in a fully-learned architecture, which does not have explicit forward-operator components. 
Despite the advantages presented, our networks still ignore the properties of the sinogram manifold, allowing for some artifacts in the resulting projections. This could explain the discontinuities in the iodine sinogram in Fig.~\ref{fig:sino}(e) and (f). In future work, additional prior information about the sinogram space could also be considered in the design of the architecture to account for its mathematical properties and reduce these artifacts. In any case, they did not have a very negative impact in the resulting image after FBP, as we show in the following results in Fig. \ref{fig:image}.

Let us compare the performance for the three different materials, as summarized in Table~\ref{tab:2}. We can see that, the fully-learned U-Net estimates the soft-tissue very poorly. This is also evident by visually inspecting Fig.~\ref{fig:image}(c), where the mono-image is blurry.
In this figure we can also appreciate the enhancement provided by the learned gradient-descent: comparing Fig.~\ref{fig:image}(d) and \ref{fig:image}(e), in \ref{fig:image}(d) the learned post-processing presents a slightly more blurred image than in \ref{fig:image}(e), which is the best among the presented results.
On the other hand, the fully-learned U-Net has a good performance in bone and iodine, i.e., the more ``sparse" channels. In fact, even though the proposed learned gradient-descent achieves the best overall performance, from observing Table~\ref{tab:2} we see that, the fully-learned network has better achievement for iodine (MSE-i) when training with 200 samples. One possible way to improve this could be to enhance the training: deep unrolled methods may present more complicated training processes, with occasional undesired events, such as vanishing gradients. Different strategies could still be tried, such as a good weight initialization, a batch normalization or a periodic learning restart, which have been shown to boost the performance of unrolled networks in CT applications \cite{Adler17PD}  \cite{Genzel21}. Nevertheless, our proposed networks are best when only 50 training samples are available, demonstrating again the robustness for a small-sample training .



Let us finally discuss the image domain metrics in Table~\ref{tab:3}. The difference between deep learning and model-based approaches is also significant in the image domain, despite the fact that FBP may add artifacts to the reconstructed images and potentially increase the noise level. It is also interesting to notice that, the proposed methods show good image results regardless of the sinogram artifacts described in Fig. \ref{fig:sino}(e) and (f). Also, these neural networks were trained with sinogram-domain metrics only, which means that in future work it would be insightful to also consider image-domain quality metrics in the training loss.



\section{Conclusion} \label{sec:Con}
We have proposed deep learning to solve sinogram-domain three-material decomposition in PCCT. We have combined the benefits of typically used model-based approaches (that lack good regularization) and fully data-driven networks (that need very big amounts of training data). Two proposed deep unrolled networks have been discussed: a learned post-processing and a learned gradient-descent. Both mimic an iterative algorithm, but the latest also contains explicit structure from the forward operator and the noise model. The proposed learned gradient-descent approach has also demonstrated the best MSE and SSIM accuracy, with fewer learned network parameters than a fully-learned U-Net, as well as more robustness to a sample-poor training. 

In conclusion, a learned gradient-descent is a very promising solution to the ill-posed non-linear three-material decomposition: it is a problem-adapted architecture that results in material-basis sinograms with less noise and less material cross-contamination. We believe that, an unrolled deep network is a very suitable alternative to the most frequently used model-based material decomposition approaches. There is, however, future research work pending to improve the training process, and also to consider a forward model that accounts for the structure of the sinogram. Also, additional testing needs to be achieved to validate its performance on more realistic data with higher resolution. Working on these future steps could be more fruitful than, instead, on improving the regularization in a model-based approach, which may result in either a very slow and complex iterative process or a still very noisy and cross-contaminated three-bases result.

\bibliographystyle{IEEEbib}
\bibliography{references}.

\end{document}